\documentclass[12pt]{article}
\usepackage[margin=1.5in]{geometry}

\usepackage{times}
\usepackage{graphicx,amsmath,amssymb,ulem,fullpage,float,tabularx,color,epsfig}
\usepackage[font=small]{caption}
\newcommand{\etal}{\textit{et. al.}}
\usepackage{titlesec}
\titleformat{\section}
  {\normalfont\normalsize}
  {\thesection}{1em}{}
\titleformat{\subsection}
  {\normalfont\small}
  {\thesubsection}{1em}{}

\title{\bf \large Developing an unfolding-incorporated coarse-grained polymer model for fibrinogen to study the mechanical behaviour}
\author{\normalsize Vivek Sharma$^{1}$ and Poulomi Sadhukhan$^{1*}$\\
\small$^1$ Bennett University, TechZone 2, Greater Noida, Uttar Pradesh 201310, India\\
\small $^*$ poulomi.sadhukhan@bennett.edu.in}

\date{}

\begin{document}

\maketitle

\thispagestyle{empty}

\begin{center}Abstract\end{center}
Fibrinogen is a protein found in blood that forms Fibrin polymer network to build a clot during wound healing process when there is a cut in the blood vessel. The fibrin fiber is highly stretchable and shows a complex mechanical properties. The fibrin monomer, Fibrinogen, has a very complex structure which is responsible for its unusual elastic behaviour. In this work, we focus on mechanism of unfolding of D-domain of Fibrinogen, and study its effect in the mechanical behaviour. We develop a coarse-grained (CG) bead-spring model for Fibrinogen which captures the unfolding of folded D-domains along with other necessary structural properties which affect the mechanical behaviour. The results from our unfolding-incorporated coarse-grained polymer (UCGP) model matches with the experimental results. This model has capacity to serve as the minimal unit to build a large-scale hierarchical structure of fibrin fiber and network to possibly unfold the mystery of fibrin's unusual elastic behaviour. This model can also be used for other polymers having folded domains or sacrificial bonds.

\section{Introduction}

Fibrinogen, a monomeric unit of fibrin fiber is found in blood which are essential element for blood clotting. When there is a wound in blood vessel and blood flows out, fibrinogen gets activated via thrombin and forms a network-like structure of fibrin fiber to stop blood flow. It is observed experimentally that the fibrin fiber can be stretched upto $3.5$ times  \cite{liu06} its original length before it breaks. Not only the unusual stretchability, it also shows peculiar elastic behaviour with multiple regimes in stress-strain behaviour. Fibrin gel can sustain a stress of the order of $10^4$ Pa upon shearing \cite{iza10} and $10$ MPa stress before it ruptures upon stretching fibrin fiber \cite{mak21}. Due to its complex elastic property, durability and easy availability, a lot of research is being done in recent years to understand the mechanism behind its unusual behaviour, which can enable us to engineer fibrin gel for desired medical use and important scaffolding material for tissue engineering \cite{li15,bar11,roja22,sanz23}. There are plenty of experimental studies on elastic properties of fibrin at fiber and network level. All-atom simulations  \cite{zhm11} have been done to mimic fibrinogen behaviour which resembles experimental data pretty well for large range of stress values showing peaks for the unfolding of D-domains of fibrinogen. But the challenges come when one wants to build a fiber or network which includes all details of monomer's force response mechanism. The all-atom simulations cannot be  extended to even build fiber  as it becomes extremely heavy computationally. Therefore, it is necessary to coarse-grain the molecular structure to some extent to develop a usable model of fiber or network. Again, the stretching experiment at fiber or network level doesn't give any clue on mechanism at the molecular level. Here comes the gap in between the information we use, at different length-scales, to model the fibrin gel in order to explain the elastic behaviour. This is why, despite availability of experimental observations of molecular composition of fibrinogen through advanced experiments, our understanding on fibrin mechanics remain poor.
We aim to develop a CG model which can be used at different length scales in order to identify which mechanism is dominating at different regimes of stress.

\begin{figure}[ht]
\centering
\includegraphics[width=10cm]{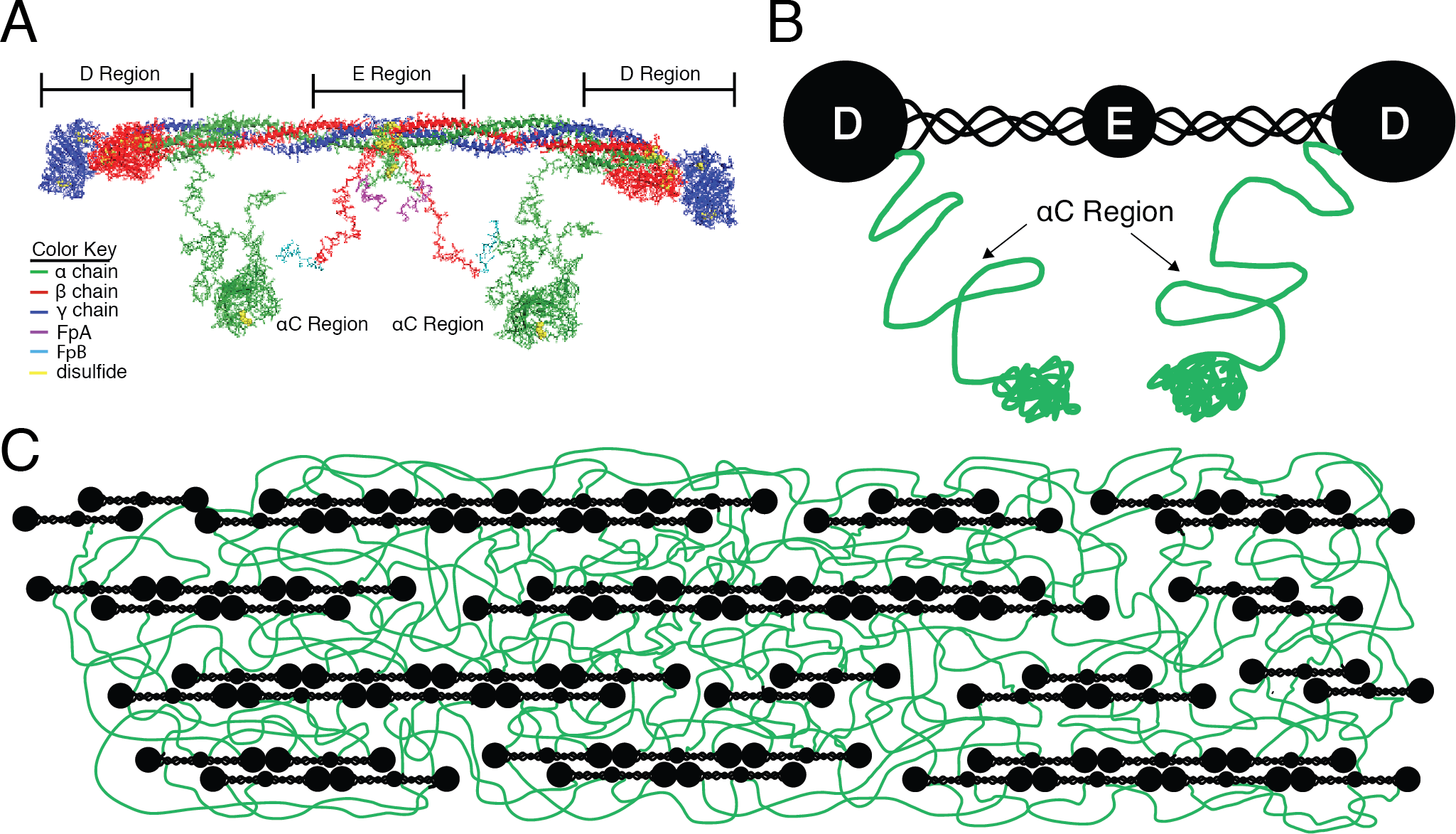}
\caption{Schematic diagram of multiscale arrangement of fibrin  \cite{buc15}. (a) Fibrinogen, a molecular unit of fibrin. (b) Schematic diagram of a molecule.  (c) Lateral half-staggered arrangement of fibrinogens and aggregation forming fiber through crosslinking via $\alpha C$ regions.}
\label{fig:structure}
\end{figure}

A fibrinogen molecule is a $340$ kDa, $45.5$  nm long complex, fibrous glycoprotein having two identical molecular halves. Each half consists of two sets of three polypeptide chains: $A\alpha$, $B\beta$, and $\gamma$ held together by multiple disulphide bonds. The molecule  is made up with three  globular domains, two outer D domains and a central E domain,  connected by two coiled-coil segments.  During thrombosis or hemostasis, through the activation by thrombin, fibrinogen starts polymerizing into protofibrils. The tripeptide globular segments are exposed to the amino-terminus of each alpha-chain residing at the center of the E-domain and combines with its complementary binding site  residing in the $\gamma$-chain in the outer D domain of another molecule. Fibrin monomer polymerizes predominantly through this `knob-hole' interactions into dimers, oligomers and then half molecule-staggered, double-stranded fibrin protofibrils. The double-stranded protofibrils then crosslink with each other to form long fiber and network. Molecular level details and mechanism related to fibrin polymerization can be obtained in the literature  \cite{brown14,sanz23,Mose05,Dool84,zhm18}. The schematic diagrams of fibrinogen, protofibril, fibrin fiber and network have been shown in Fig.\ref{fig:structure}  \cite{buc15}

A typical fibrin clot can be seen in scanning electron microscopy (SEM) and confocal fluorescence microscopy  \cite{SEM, cagny16}, which shows the arrangement of fibers of different thickness into a network. A comparison of images before and after stretching gives us idea about how stretching affect the fiber alignment. An interesting mechanical behaviour is observed for fibrin gel in rheology experiments which is done for wide value of shear strain, upto $100$\%   \cite{iza10}. The tangential modulus vs stress graph shows five distinct regimes, among which first regime (due to entropic stretching of fibers) and fifth regime (backbone stretching showing $\sigma^{3/2}$ dependence) are usually observed for any biogels. The entropic regime is predicted from any model of biopolymers, and the $\sigma^{3/2}$ behaviour is obtained for inextensible polymers and can be predicted as strain-stiffening behaviour of worm-like chain. An additional intermediate plateau region appears for fibrin gel and it is also known for its high stretchability. So, before it reaches its inextensible regimes, the fiber must go through the condition where the fibrinogen unfolds to accommodate the build up stress. Therefore, we speculate that possibly unfolding is responsible for the intermediate nonlinear regime.

The fiber stretching experiments done by using atomic force microscopy (AFM) to understand the mechanical behaviour before the fiber ruptures at around $212$\% strain or above $10$ MPa stress  \cite{mak21, liu10}. It shows mostly bi-linear behaviour, where initial part is entropic and later part is enthalpic. This is a very generic which most polymers show upon stretching. So the AFM stretching experiments have not given any unusual results for fiber yet, so that we can guess on at which point unfolding may play role when we stretch fibrin polymer. In order to get the clear picture on mechanism of domain unfolding of fibrinogen, the obvious option would be an fibrinogen stretching experiment. AFM stretching experiments have been done for single fibrinogen and protofibril  \cite{lim08, ave08}. Zhmurov {\it \etal}  \cite{zhm11} have done fibrinogen pulling simulation for a single $\gamma$-nodule, and half fibrinogen molecule including stretching of $\beta$-nodule, coiled-coil part and central E domain. The other mentioned regions gets stretched but sustain the force, while the $\gamma$-nodule gets unfolded in three steps, indicated by a dip in the force-extension curve for every unfolding. An all-atom simulation is also done by Zhmurov \etal
 and estimated the peak force before unfolding and the elongation of molecule.

Apart from the all-atom simulations, one needs to do coarse-grain simulation also to scale up the system. The mechanical bead-spring model has been adopted for simulating fiber and networks to see the effect of crosslinking and unfolding. In Ref. \cite{fila23}, they have developed a model of fiber by using a bead-spring system with effective spring constant of a fiber. The spring constant and the slopes of the bi-linear stretching curve have been evaluated by comparing with the experimental curves of fiber stretching, and then related the values with standard polymer parameters like persistence length, stretching and bending stiffness etc. Similar work has been done by Maksudov \etal \cite{mak21}. To capture the nonlinearity into the theoretical model, Yesudasan \etal  \cite{yesu20} has developed a spring network with variable spring stiffness. The spring stiffness changes the value once the extension crosses certain value, and the linear spring becomes a cubic spring. Due to the introduction of nonlinear terms the force-extension curve becomes nonlinear. They have estimated the network force and elasticity from 8-chain network model. The study on the effect of unfolding in force-extension behaviour of fiber is lacking. This the main focus of this work.
In section \ref{sec:model}, we describe the coarse-grained model that we develop and discuss the force-extension behaviour obtained from our model. We conclude in section \ref{sec:conclude} and discuss about the scope of using this model for other polymers which exhibit domain unfolding or sacrificial bonds.

\section{Model And Results}\label{sec:model}
To build a coarse-grained bead-spring model, which can include unfolding of $\gamma$-nodule, we modified the nine-bead model used for fibrin polymerization described in Yesudasan \etal \cite{yesu18}. To accommodate unfolding and staggered packing, we split the D and E regions into three beads connected by two springs representing $\beta$ and $\gamma$-nodules.
The springs are harmonic having spring energy $E_s=k (r-r_0)^2$, where $k$
is spring stiffness, $r_0$ is the rest length, different for different type of spings. The bending stiffness is introduced by harmonic term $E_b=k_\theta (\theta-\theta_0)^2$, where $\theta_0=180^o$.  The ``soft'' potential has been used as non-bonded soft-core interaction. Left end of the bead-spring chain has been anchored to wall, whereas the right end of the chain is pulled at constant speed in order to apply force on the chain. The whole dynamics is studied at $T=300$K by integrating Langevin equation in LAMMPS  \cite{lammps}.
Table \ref{tab1} summarizes the lengths of each springs and corresponding spring constants before and after unfolding.

\begin{figure}[t]
\centering
 \includegraphics[width=10cm]{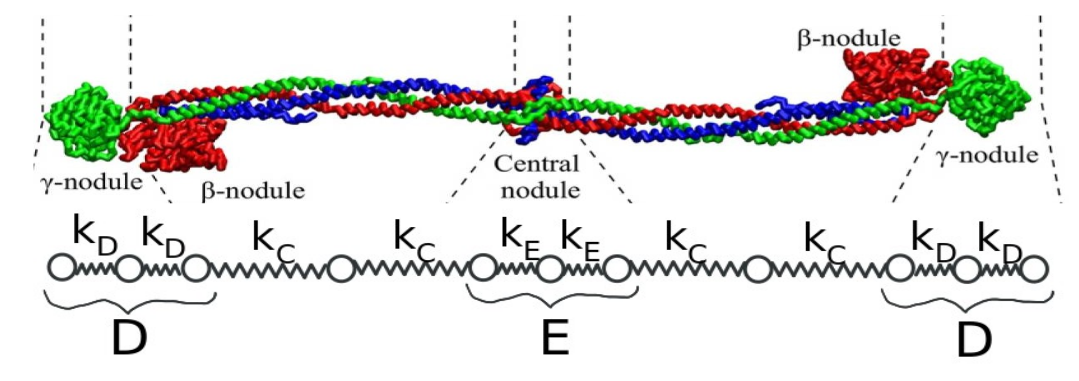}
\caption{Schematic diagram of the equivalent bead-spring model (bottom) consisting of $11$ beads. The $\gamma$-nodule in D region and central E regions are built with three beads connected with springs of stiffness $k_D$ and $k_E$. The coiled-coil part is built with two springs of stiffness $k_C$. (Spring lengths are not to the scales in the figure.)}
\end{figure}

In fibrinogen, each D domain measures 5 nm, the E domain measures 1.5 nm, and each coiled-coil segment measures 17 nm, totalling to  $45.5$  nm of fibrinogen  \cite{zhm11}. As each of D, E and coiled-coil part is built with two harmonic springs in our model, initial rest-lengths of each springs are taken as 2.5 nm, 0.75 nm and 8.5 nm, respectively. After three steps of unfolding of each D domain, the length of the each unfolded part becomes 80 nm. The initial values of spring stiffness $k_C$, $k_D$ and $k_E$ are calculated from the slope of force-extension graph of AFM experiment \cite{zhm11}. The calculated ratio of spring stiffnesses comes out to be 1:3:10 for $k_c$, $k_D$ and $k_E$, based on the rest-lengths mentioned above.

The pulling simulation has been done with pulling speed 10$\mu$m/s.
As the unfolding happens in three steps, we set three threshold values of stress, attaining which the $\gamma$-nodule undergoes unfolding by changing rest-length and stiffness.
At the first unfolding the rest-length is changed to $r_0\to r_0^{(1)}=a_0 r_c$, where $r_c$ is the current length of the spring and the factor $a_0$ can take value $0.6-1$. At this point, we change the spring constant from $k_D$ to $k_D'$ which is $10$  pN/nm. The remaining two unfolding happen at the higher stress threshold
% (1500 and 3000 in LAMMPS LJ unit)
with same update condition of rest-length, $r_0^{(i)}\to r_0^{(i+1)}=a_0 r_0^{(i)}$ ($i=2,3$) keeping the  spring stiffness same.

\begin{table}[hb]
\centering
\begin{tabular}{cccc}
\hline
Quantities & Coiled-coil & E region & D region \\
\hline
Stiffness & 11  pN/nm & 110  pN/nm & 33  pN/nm $\to$ 10 pN/nm\\
Length & 8.5 nm & 0.75 nm & 2.5 nm \\
\hline
\end{tabular}
\caption{The details of spring connectors of each region, spring stiffness, rest lengths before and after unfolding.}
\label{tab1}
\end{table}

The tuning the parameters, (i) spring stiffness and (ii) rest-length update conditions at each unfolding, and (iii) threshold stress values, have to be determined in such a way that we get six peaks for two folded domains and the unfolding process completes by $150$ pN force on molecule. Therefore, the values mentioned here can be tuned in a range to achieve the desired output. The updated $k_D'$ value controls the overall slope of the force-extension graph, while the initial $k_D$ value controls the slope till first unfolding.

\begin{figure}[t]
\centering
\includegraphics[width=8.2cm,clip]{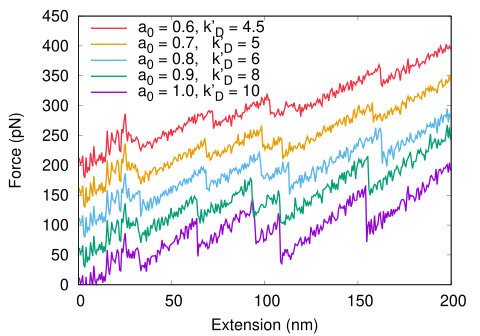}
\includegraphics[width=8.2cm,clip]{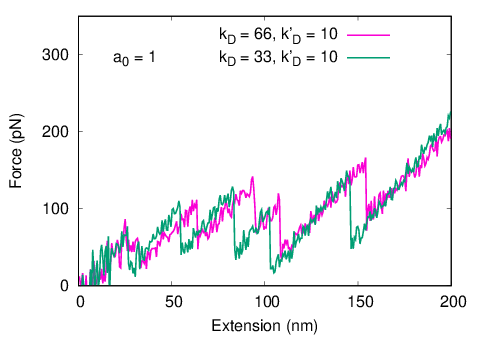}
\caption{Force-extension curve obtained from the simulation. (Left) Different curves show how the behaviour changes with the choice of $a_0$. For different values of $a_0$,  $k_D'$ is adjusted to get correct slope and peak position with $k_D=33$  pN/nm.
(Right) Comparison of force-extension behaviour for different set of spring stiffnesses of $k_D=33$  pN/nm and $k_D=66$  pN/nm, with $a_0=1$.}
\label{fig:f-x}
\end{figure}

The force-extension graph is shown in Fig. \ref{fig:f-x} (Left) for various choices of $a_0$, where $a_0$ is the fraction of current length the spring has at a certain value of force. For better representation of graph, we have shifted $y$ offset by 50 unit to avoid overlap. Each curve is showing six peaks indicating six unfolding events occurring in two D domains. This graphs matches the all-atom simulation results mimicking the unfolding of fibrinogen in silico presented in Ref. \cite{zhm11}. Higher the value of $a_0$, more relaxed the spring becomes after unfolding. Therefore, the drop in force value is higher at each unfolding. This requires the control on $k_D'$ value as well to maintain the overall slope of the curve. From our simulation, the peak positions are also coming approximately in the desired values of extension, the first and last ones being at around $30$ and $150$  nm. In the Fig.\ref{fig:f-x} (Right), we compare how the force-extension behaviour changes as we change the spring stiffnesses keeping the ratio as mentioned in Sec.\ref{sec:model}. The peak positions changes slightly but the overall slope does not change, as the slope primarily depends on the updated spring constant corresponding to D-domain. The peak positions can be adjusted by tuning the stress threshold values. As the fibrinogen molecule does not behave as a worm-like chain (WLC)  \cite{yesu20},
% Hence, we should not expect WLC-like behaviour in
the segment of the curve in between two peaks are expected to be linear.
The curvature may appear in between initial few peaks due to the entropic nature of molecule chain.

\section{Conclusion}\label{sec:conclude}

The fibrin polymer has its own structural complexity, and therefore, exhibit unusual mechanical response to the external stress. In spite of fibrin's abundance in nature and handful of available experimental observations, there exists lack of understanding of the stress response mechanism taking place in the different regime of stress on fibrin network. In order to investigate the possible reason of complex stress-strain behaviour, we have proposed an unfolding-incorporated coarse-grained polymer (UCGP) model for fibrinogen that provides the important structural characteristics and molecular mechanism during stretching of fibrinogen through step-wise unfolding of $\gamma$-nodule. This minimal bead-spring model captures the unfolding through the update in spring constant and rest-length upon building up a threshold stress on a spring representing the $\gamma$ chain. A three-step unfolding of each $\gamma$-nodule exhibits the behaviour as seen in AFM experiment and all-atom simulation  \cite{zhm11}.

In this paper, we have focused exclusively on the development of a coarse-grained bead-spring model which would be just enough to produce the results of all-atom stimulation and AFM experiment \cite{zhm11}. This model indeed has the scope to extend to fiber level by considering the CG fibrinogen unit as the building block of fiber. The multiscale modeling discusses upscaling of molecular-level behavior to larger structures and how molecular interactions give rise to macroscopic properties. This can be achieved from our model by introducing 'knob-hole' interactions and cross-linking of fibrin represented as nodes and edges. As the knob-hole interaction is very strong, we can model it by taking it as a spring with a high spring constant and a very small rest length. In the case of $\alpha$C crosslinking, several articles suggest the effect and structural information about the $\alpha$C \cite{zhm18}. We speculate that the extra length of fiber upon unfolding of domains, which we introduced in UCGP model, may show plateau-like behaviour in the elastic modulus. In fact, our primary results of fibers and networks based on current model shows the expected behaviour as seen in experiments.

This model is the first of its kind that incorporates the domain unfolding through bead-spring model which can be used to build multiscale structures. Not only this model captures the fibrinogen domain unfolding, but also can be applied to the polymers with sacrificial bonds, which shows similar force-extension curve  \cite{fant06,zhou17}, by changing the update criteria of rest-length upon breaking the sacrificial bonds. Therefore, this model can be very useful to have insight of the mechanism at the microscopic level for wide range of applied stress on fibrin, and open the avenue of new studies on designing polymer fibers and networks with rich elastic properties.


\begin{thebibliography}{25}
\bibitem{liu06} Liu, W. Jawerth, LM \etal (2006) Fibrin fibers have extraordinary extensibility and elasticity, Science 313, 634-634.
%
\bibitem{iza10} Piechocka, IK; Bacabac, RG \etal (2010) Structural Hierarchy Governs Fibrin Gel Mechanics, Biophys J. 98 (10).
%
\bibitem{mak21} Maksudov, F; Daraei, A \etal (2021) Strength, deformability and toughness of uncrosslinked fibrin fibers from theoretical reconstruction of stress-strain curves, Acta Biomater. 136, 327-342.
%
\bibitem{li15} Li, Y; Meng, H \etal (2015) ibrin gel as an injectable biodegradable scaffold and cell carrier for tissue engineering. ScientificWorldJournal, 685690. doi: 10.1155/2015/685690.
%
\bibitem{bar11} Barsotti, MC; Felice, F \etal (2011)  Fibrin as a scaffold for cardiac tissue engineering,  58 (5) 301-310.
%
\bibitem{roja22} Rojas-Murillo, JA; Simental-Mendia, MA \etal (2022) Physical, Mechanical, and Biological Properties of Fibrin Scaffolds for Cartilage Repair, Int. J  Mol. Sci. 23 (17) 9879.
%
\bibitem{zhm11} Zhmurov, A; Brown, AEX \etal (2011) Mechanism of Fibrin(ogen) forced unfolding, Structure 19, 1615-1624.
%
\bibitem{brown14} Brown, AC; Barker, TH (2014) Fibrin-based biomaterials: modulation of macroscopic properties through rational design at the molecular level, Acta Biomater. 10 (4): 1502-14.
%
\bibitem{sanz23}
Sanz-Horta, R; Matesanz, A \etal (2023) Technological advances in fibrin for tissue engineering, Journal of Tissue Engineering. 14.
%
\bibitem{Mose05} Moseson, MW; (2005) Fibrinogen and fibrin structure and functions, J Thromb Haemost. 3 (8) 1894.
%
\bibitem{Dool84} Doolittle, RF (1984)  Fibrinogen and fibrin, Annu Rev Biochem. 53, 195-229.
%
\bibitem{zhm18} Zhmurov, A; Protopopova, AD \etal (2018) Atomic Structural Models of Fibrin Oligomers, Structure 26 (6) 857-868.
%
\bibitem{buc15} Bucay, I; O’Brien, ET III \etal
(2015) Physical Determinants of Fibrinolysis in Single Fibrin Fibers, PLoS ONE 10(2): e0116350.
%
\bibitem{SEM} Janmey, PA; Winer, JP \etal (2009) Fibrin gels and their clinical and bioengineering applications,  J R Soc. Interface 6 (30) 1-10.
%
\bibitem{cagny16} de Cagney, HCG; Vos, BE \etal  (2016)  Porosity Governs Normal Stresses in Polymer Gels, Phys. Rev. Lett. 117, 217802.
%
\bibitem{liu10} Liu, W; Carlisle, CR \etal (2010) The mechanical properties of single fibrin fibers, Journal of Thrombosis and Haemostasis 8 (5).
%
\bibitem{lim08} Lim, BBC; Lee, EH \etal (2007) Molecular Basis of Fibrin Clot Elasticity, Structure 16, 449-459.
%
\bibitem{ave08} Averett, LE; Geer, CB \etal (2008) Complexity of ``A-a'' Knob-Hole Fibrin Interaction Revealed by Atomic Force Spectroscopy, Langmuir 24, 9, 4421-5196.
%
\bibitem{fila23} Filla, N; Zhao, Y and  Wang, X (2023) Fibrin fiber deformation mechanisms: insights from phenomenological modeling to molecular details, Biomech Model Mechanobiol 22, 851-869.
%
\bibitem{yesu20} Yesudasan, S and Averett, RD (2020) Multiscale Network Modeling of Fibrin Fibers and Fibrin Clots with Protofibril Binding Mechanics,  Polymer 12 (6).
%
\bibitem{yesu18} Yesudasan, S; Averett, RD (2018) Coarse-grain molecular dynamics simulation of fibrin polymerization, J Mol.  Model 24, 109.
%
\bibitem{lammps} LAMMPS - a flexible simulation tool for particle-based materials modeling at the atomic, meso, and continuum scales, AP Thompson \etal Computer Physics Communications (2022) 271, 108171. Website: https://www.lammps.org/index.html.
%
\bibitem{fant06} Fantner, GE; Oroudjev, E \etal (2006) Sacrificial Bonds and Hidden Length: Unraveling Molecular Mesostructures in Tough Materials, Biophys. J 90 (4) 1411-1418.
%
\bibitem{zhou17} Zhou, X; Guo, B \etal (2017) Progress in bio-inspired sacrificial bonds in artificial polymeric materials,  Chem. Soc. Rev., 46, 6301.

% \bibitem{wies05}
% % The elasticity of individual fibrin fiber in a clot,
% JP Collet \etal, (2005) PNAS 102 (26) 9133-9137. % for typical fibrin length range


% \bibitem{SW99} Strunk Jr., W., \& White, E. B. (1999). {\it The Elements of Style, 4th Edition.} Pearson.
\end{thebibliography}
\end{document}